\documentclass[letterpaper, 10 pt, conference, dvipsnames]{ieeeconf}
\IEEEoverridecommandlockouts                            
\title{\LARGE \bf Estimates on the domain of validity for Lyapunov-Schmidt reduction}

\author{Pranav~Gupta\(^{1}\),
        Anastasia~Bizyaeva\(^{2}\),
        and~Ravi~Banavar\(^{3}\)
    \thanks{\(^1\) Mechanical Engineering, Indian Institute of Technology Bombay, Mumbai, 400076, Maharashtra, India {\tt\small guptapranav@iitb.ac.in}}
    \thanks{\(^{2}\) Sibley School of Mechanical and Aerospace Engineering, Cornell University, Ithaca NY, USA, 14853 {\tt\small anastasiab@cornell.edu}}
    \thanks{\(^{3}\) Systems and Control Engineering, Indian Institute of Technology Bombay, Mumbai, 400076, Maharashtra, India {\tt\small banavar@iitb.ac.in}}
}
\usepackage{mymacros}

\begin{document}

\maketitle
\thispagestyle{empty}
\pagestyle{empty}


\begin{abstract}
    Lyapunov-Schmidt reduction is a dimensionality reduction technique in nonlinear systems analysis that is commonly utilised in the study of bifurcation problems in high-dimensional systems. The method is a systematic procedure for reducing the dimensionality of systems of algebraic equations that have singular points, preserving essential features of their solution sets. In this article, we establish estimates for the region of validity of the reduction by leveraging recently derived bounds on the {Implicit Function Theorem}. We then apply these bounds to an illustrative example of a two-dimensional system with a pitchfork bifurcation.
\end{abstract}


\section{Introduction}

A bifurcation in a parameterised nonlinear dynamical system marks a point at which the system undergoes qualitative changes in behaviour, often leading to the creation of multiple new isolated equilibria, onset of oscillation, and chaotic dynamics \cite{guckenheimer2013nonlinear}. Analysis of bifurcations plays an important role in both \textit{understanding} and \textit{controlling} systems with complex nonlinear features including power systems \cite{ajjarapu1992bifurcation}, epidemic processes \cite{pare2020modeling}, social dynamics \cite{bizyaeva2022nonlinear,bizyaeva2023thesis}, and neuromorphic computing networks \cite{juarez2024analysis}. In a control context, objectives range from understanding when feedback laws give rise to bifurcations and potentially suppressing their onset  \cite{chen1994overview,chen2000bifurcation}, to designing dynamics and feedback laws that give rise to meaningful behaviour through a bifurcation \cite{leonard2024fast}.

Dimensionality reduction is a key step in formally classifying a bifurcation. While a system of interest may have high state dimension, bifurcations typically occur along low-dimensional manifolds. Identifying low-dimensional representations of systems that preserve their essential features on or near these manifolds allows for a more tractable investigation of dynamics near bifurcation points. Centre manifold reduction and Lyapunov-Schmidt reduction are two methods commonly utilised for this purpose.

The centre manifold reduction method is used to simultaneously approximate a nonlinear manifold of a bifurcation along with a low-dimensional representation of the dynamical system with respect to coordinates along this manifold \cite[Chapter 3.2]{guckenheimer2013nonlinear}. While this technique is powerful, it requires performing calculations that are challenging to carry out for systems with high state dimension and complex structure such as large-scale networks. For such systems, Lyapunov-Schmidt reduction is often a more appropriate tool
\cite{sidorov2013lyapunov,golubitsky2023dynamics}.

Lyapunov-Schmidt reduction is distinguished from centre manifold reduction by its topological viewpoint focused on studying properties of parameterised curves of equilibria. The end result of performing Lyapunov-Schmidt reduction is a low-dimensional parameterised algebraic system of equations whose zeros are in one-to-one correspondence with equilibria of the original dynamical system near its bifurcation. This allows for local bifurcation analysis without the need to derive explicit approximations of a nonlinear manifold, for which tools from singularity theory can be utilised \cite{Golubitsky1985}.

A notable limitation of Lyapunov-Schmidt reduction is the local nature of its results, which hold on arbitrarily small neighbourhoods of a system's bifurcation point. This limitation can be traced to a key step in the reduction procedure being an application of the Implicit Function Theorem (ImFT), whose conclusions are local. Locality is often not an issue when one's goal is to simply classify the nature of a bifurcation. However, for many engineering applications it is also important to establish explicit bounds of validity for the analysis, for example in order to quantify robustness of the established features to parameter uncertainty. {While validity bounds have been explored in the case of centre manifold reduction \cite{roberts2018backwards}, analogous bounds for Lyapunov-Schmidt reduction have not been established to the authors' best knowledge. In this paper we derive validity bound estimates for Lyapunov-Schmidt reduction, leveraging} recent advances in estimating the domain of validity for the ImFT \cite{jindal2023imft,jindal2023fdbklin}.

The primary contributions of the presented work are 1) a derivation of bounds of validity for the Lyapunov-Schmidt reduction procedure for finite-dimensional systems of nonlinear algebraic equations, and 2) detailed computation of these bounds for a low-dimensional illustrative example. 

The paper is organised the following way. In section \ref{sec:prelims} we introduce notation and background information on the Implicit Function Theorem and the Lyapunov-Schmidt reduction procedure. Then in section \ref{sec:bounds} we establish our main result and derive the bounds. We apply these bounds to a low-dimensional illustrative example in section \ref{sec:example} and conclude in section \ref{sec:conclusion}.

\section{Preliminaries \label{sec:prelims}}

\subsection{Notation and definitions}

Let \(\R\) denote the set of real numbers and \(\mathbb{N}\)  denote the set of positive integers.
Throughout the text we let \(U \subseteq \R[n]\) and \(V \subseteq \R[m]\) denote open subsets of Euclidean spaces with appropriate dimensions \(m,n\).
We will say a map \(\mathbf{f}:U \to V\) belongs to class \(C^\nu\) for \(\nu\in\mathbb{N}\) if it is \(\nu\)-times continuously differentiable.
An open ball of radius \(r>0\) centred at \(\mathbf{x}_0 \in \R[n]\) is denoted by \(\mathcal{B}(\mathbf{x}_0,r) = \{\mathbf{x}\in\R[n]:\norm{\mathbf{x}-\mathbf{x}_0}<r\}\) where \(\norm{\cdot}\) is a vector norm on \(\R[n]\).
For any matrix \(A\in\R[s \times t]\) we use \(\norm{A}\) to denote a matrix norm induced by the same vector norm, i.e., \(\norm{A} := \sup\{\norm{A\mathbf{x}}:\norm{\mathbf{x}}=1\}\).
For a \(f \in \mathcal{C}^{\nu}\), \(Df\) denotes the Jacobian matrix and \(\mathrm{D}_{\mathbf{y}} f\) denotes a Jacobian matrix of partial derivatives with respect to the variables \(\mathbf{y}\). \({\mathcal{N}}(\mathbf{x})\) denotes an open neighbourhood of \(\mathbf{x}\).{ We use the notation $\tanh(\mathbf{x}) = \begin{bmatrix} \tanh(x_1) & \dots & \tanh(x_n) \end{bmatrix}^\top$ for $\mathbf{x} \in \R[n]$.} 

\subsection{Implicit Function Theorem}

Consider a \(\mathcal{C}^{\nu}\) map \(f: U \times V \to \R[m]\) with \(\nu \geq 1\).
The {Implicit Function Theorem} is a fundamental result in mathematical analysis that establishes conditions under which the variables \(\mathbf{x} \in U\) implicitly define the variables \(\mathbf{y} \in V\) through the relation \(f(\mathbf{x},\mathbf{y}) = w_0\) for \(w_0 \in \R\) \cite{krantz2002implicit}. We first state a version of this standard result from \cite[Theorem 4.B]{zeidler2013vol1}.

\thm[Implicit Function Theorem] \label{thm:ImFT} Let \(U\subset\R[n]\) and \(V\subset\R[m]\) be be open sets. Consider {a} \(\mathcal{C}^{\nu}\) smooth map
\begin{equation}
    U\times V\ni(\mathbf{x},\mathbf{y})\mapsto f(\mathbf{x},\mathbf{y})\in\R[m] \label{eq:imftmap}
\end{equation}
where \(\nu\geq1\), {and consider} a point \((\mathbf{x}_0,\mathbf{y}_0)\in U\times V\) for which \(\mathrm{D}_{\mathbf{y}}f(\mathbf{x}_0,\mathbf{y}_0):\R[m]\to\R[m]\) is an isomorphism. Then for any \({\mathcal{N}}(\mathbf{y}_0)\subset V\), there exists an \({\mathcal{N}}(\mathbf{x}_0)\subset U\) and a {unique} \(\mathcal{C}^{\nu}\) map \(g:{\mathcal{N}}(\mathbf{x}_0)\to{\mathcal{N}}(\mathbf{y}_0)\) satisfying
\begin{equation}
    f(\mathbf{x},g(\mathbf{x})) = f(\mathbf{x_0},\mathbf{y}_0)\qquad\forall\,\mathbf{x}\in{\mathcal{N}}(\mathbf{x}_0) \label{eq:imft}
\end{equation}
{Moreover, if \(\mathrm{D}_{\mathbf{y}}f(\mathbf{x},g(\mathbf{x}))\) is invertible, we also have 
\vspace{-1.5mm}\begin{equation}
    \mathrm{D}_{\mathbf{x}}g(\mathbf{x},\mathbf{y})=-\mathrm{D}_{\mathbf{y}}f(\mathbf{x},g(\mathbf{x}))^{-1}\mathrm{D}_{\mathbf{x}}f(\mathbf{x},g(\mathbf{x}))
\end{equation}\vspace{-5.5mm}}

\begin{figure}[ht!]
    \centering
    \includegraphics[width=\linewidth]{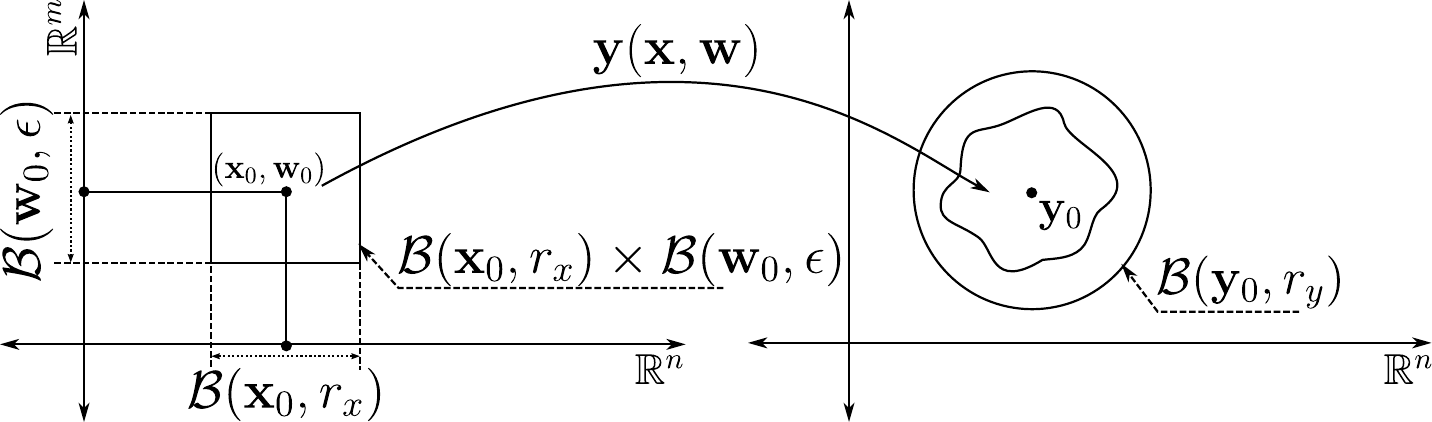}
    \caption{a schematic of the bounds on Implicit Function Theorem: \((x,w)\in\ball{x_0}{r_x}\times\ball{w_0}{\epsilon}\mapsto y(x,w)\in\ball{y_0}{r_y}\) where \(w_0 := f(x_0,y_0)\) satisfying \(f(x,y(x,w)) = w\); for our purposes we assert \(\epsilon=0\) (Fig 2. from \cite{jindal2023fdbklin}) \label{fig:ImFT}}
    \vspace{-6mm}
\end{figure}

We now state explicit bounds on the regions in which the results of the ImFT presented in \cite[Corollary 3.8]{jindal2023imft} below.
\thm\label{thm:imftbounds} Define quantities \(M_x\) and \(M_y\), dependent on the Jacobians with respect to \(\mathbf{x}\) and \(\mathbf{y}\) as:
\begin{align}
    M_x & := \norm{\mathrm{D}_{\mathbf{x}}f(\mathbf{x}_0,\mathbf{y}_0)} \label{def:M_x}                \\
    M_y & := \norm{\big(\mathrm{D}_{\mathbf{y}}f(\mathbf{x}_0,\mathbf{y}_0)\big)^{-1}} \label{def:M_y}
\end{align}
For any given \(r_x,r_y>0\), define \(\mathbf{L}_\mathbf{x}\) and \(\mathbf{L}_\mathbf{y}\) as:
\begin{align}
    \mathbf{L}_\mathbf{x}(r_x) := \sup\{     & \norm{\mathrm{D}_\mathbf{x}f(\mathbf{x},\mathbf{y}_0)-\mathrm{D}_\mathbf{x}f(\mathbf{x}_0,\mathbf{y}_0)}:   \notag \\
                                           & \mathbf{x}\in\mathcal{B}(\mathbf{x}_0,r_x)\}                                                 \label{def:L_x}       \\
    \mathbf{L}_\mathbf{y}(r_x,r_y) := \sup\{ & \norm{\mathrm{D}_\mathbf{y}f(\mathbf{x},\mathbf{y})-\mathrm{D}_\mathbf{y}f(\mathbf{x}_0,\mathbf{y}_0)}: \notag     \\
                                           & (\mathbf{x},\mathbf{y})\in\mathcal{B}(\mathbf{x}_0,r_x)\times\mathcal{B}(\mathbf{y}_0,r_y)\} \label{def:L_y}
\end{align}
Then for all \(r_x,r_y>0\) satisfying
\begin{align}
    \mathbf{L}_\mathbf{x}(r_x)r_x+\mathbf{L}_\mathbf{y}(r_x,r_y)r_y & < \frac{r_y}{M_y}-M_xr_x \label{eq:1} \\
    M_y\mathbf{L}_\mathbf{y}(r_x,r_y)                               & < 1                      \label{eq:2}
\end{align}
there exists a smooth map \(g:\mathcal{B}(\mathbf{x}_0,r_x)\to\mathcal{B}(\mathbf{y}_0,r_y)\) that satisfies \(f(\mathbf{x},g(\mathbf{x})) = f(\mathbf{x}_0,\mathbf{y}_0)\), i.e. validity of the map \(g\) obtained from the Implicit Function Theorem \ref{thm:ImFT} is guaranteed within at least the open ball neighbourhoods \({\mathcal{N}}(\mathbf{x}_0)\equiv\mathcal{B}(\mathbf{x}_0,r_x)\) as its domain and \({\mathcal{N}}(\mathbf{y}_0)\equiv\mathcal{B}(\mathbf{y}_0,r_y)\) as its co-domain. A graphical schematic of the same has been provided in Fig. \ref{fig:ImFT}

\subsection{Bifurcation problems and Lyapunov-Schmidt reduction}

Consider a parameterised nonlinear system
\begin{equation}
    \dot{\mathbf{x}} = \boldsymbol{\Phi}(\mathbf{x},\lambda) \label{eq:dyn_sys}
\end{equation}
where \(\mathbf{x} \in \R[n]\) is the system state, {\(\lambda \in \R[m]\) is a parameter vector}, and \({\boldsymbol{\Phi}}\) is at least {\(\mathcal{C}^{2}\)}. The study of \textit{local bifurcations} in \eqref{eq:dyn_sys} concerns classifying local changes in the number and stability of equilibria, periodic orbits, and other invariant sets of the system as its {parameter vector \(\lambda\)} is varied. 

Equilibria of \eqref{eq:dyn_sys} are level sets of the algebraic system of equations
\begin{equation}
    \boldsymbol{\Phi}(\mathbf{x},\lambda) = 0. \label{eq:equilibria}
\end{equation}
{A \textit{bifurcation}} \textit{point} of equilibria of \eqref{eq:dyn_sys} is a point \((\mathbf{x}_0,\lambda_0)\in\R[n]\times\R[m]\) that satisfies the following:
\begin{enumerate}
    \item The point \((\mathbf{x},\lambda) = (\mathbf{x}_0,\lambda_0)\) is a solution of \eqref{eq:equilibria};
    \item The Jacobian matrix \(J = \mathrm{D}_{\mathbf{x}}{\boldsymbol{\Phi}}(\mathbf{x}_0,\lambda_0)\) has at least one {zero} eigenvalue. \footnote{More generally, at a bifurcation point $J$ can have purely imaginary sets of eigenvalues; in this paper we focus only on bifurcations of equilibria.}
\end{enumerate}
As a consequence of the ImFT, in a neighbourhood of any hyperbolic fixed point of \eqref{eq:dyn_sys}, the parameter vector \(\lambda\) uniquely parametrizes a curve of nearby equilibria \((\mathbf{x},\lambda) = (g(\lambda),\lambda)\) where  \(g:\R[m]\to\R[n]\) is \(\mathcal{C}^2\). Intuitively, this means that the number of equilibria in a neighbourhood of a hyperbolic fixed point is preserved as \(\lambda\) is varied. Bifurcation points are then precisely the points at which the ImFT fails, potentially leading to novel features in the flow of  \eqref{eq:dyn_sys}.

It is often the case that \eqref{eq:dyn_sys} is large in state dimension. A classic result in dynamical systems theory constrains the dimension of the invariant manifold on which nontrivial features arise near bifurcation points to the number of singular eigenvalues of the linearised Jacobian \(J\) at the bifurcation point \cite[Theorem 3.2.1]{guckenheimer2013nonlinear}. In cases where  \(J\) has a single zero eigenvalue, systems of arbitrarily high state dimension are often effectively one-dimensional near a bifurcation point \((\mathbf{x}_0, \lambda_0)\). Lyapunov-Schmidt reduction is a dimensionality reduction technique that enables us to derive a low-dimensional representation of the algebraic system \eqref{eq:equilibria} that preserves the topology of the curves of equilibiria.

The Lyapunov-Schmidt reduction method is in its essence a clever application of the ImFT at points where the standard assumptions stated in Theorem \ref{thm:ImFT} are not satisfied. Where the ImFT guarantees the existence of a full-dimension local map without the need of altering coordinates, Lyapunov-Schmidt reduction provides a compromise for systems with singular points: a reduced-dimension local map expressed in a coordinate system formed from the splitting of the domain concerned into orthogonal complement subspaces. We now state the main premise of the reduction as a theorem, based on\cite[Ch. VII]{Golubitsky1985}.

\begin{figure}[ht!]
    \centering
    \includegraphics[width=\linewidth]{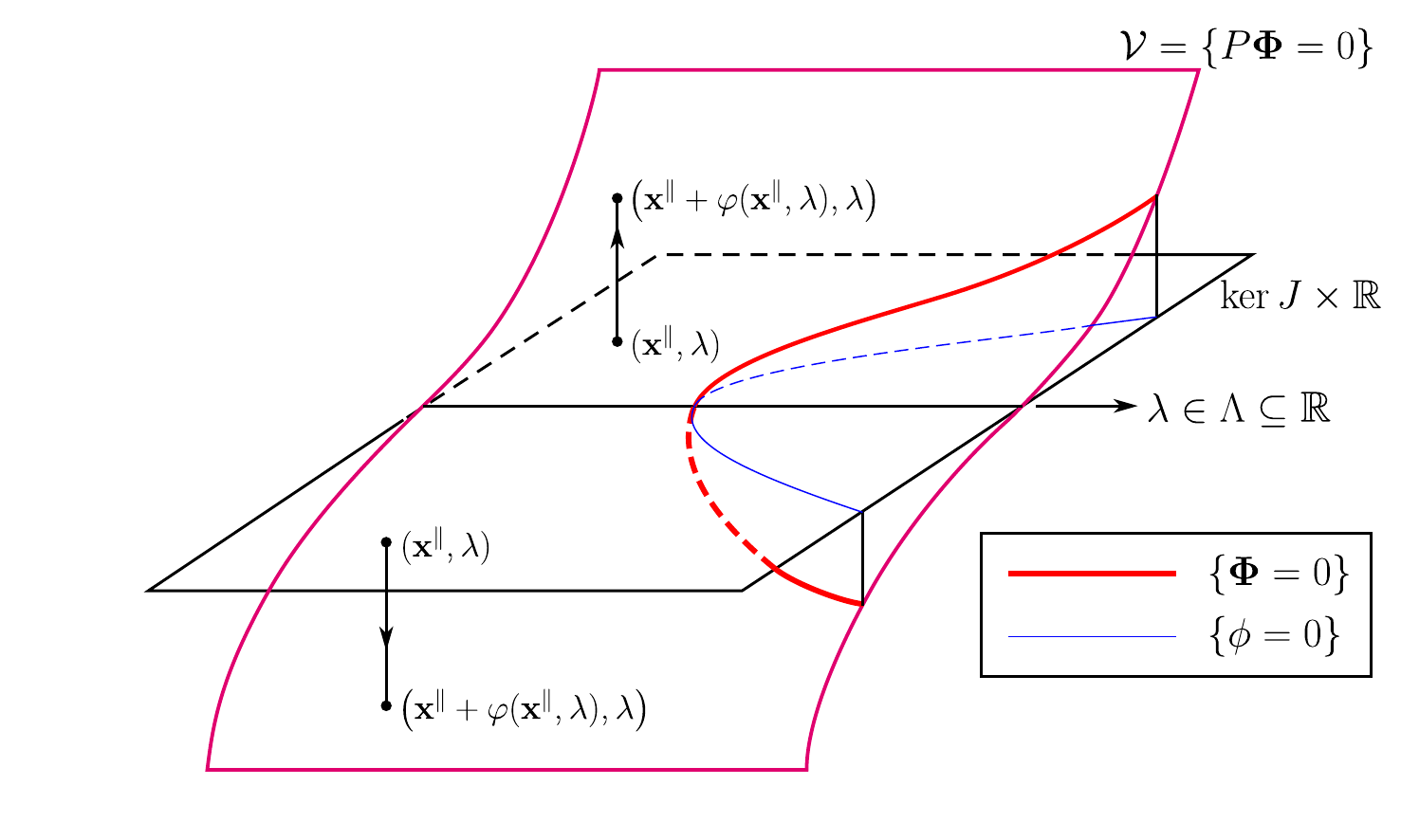}
    \caption{sketch of \(\ker(J)\times\R\) and the zero sets of \(\boldsymbol{\Phi},P\boldsymbol{\Phi}\) and \(\phi\) in for a 2-dimensional system that has a pitchfork bifurcation. Figure based on \cite[Ch.I, Fig. 3.1]{Golubitsky1985} \label{fig:pitchfork}}
\end{figure}

\thm\label{thm:LSR} Let \(X\subset\R[n]\) and \(\Lambda\subset\R[m]\) be open sets. Let \(\boldsymbol{\Phi}:X\times\Lambda\to\R[n]\) be a smooth map with \(\boldsymbol{\Phi}(\mathbf{x}_0,\lambda_0) = 0\) for some \((\mathbf{x}_0,\lambda_0)\in X\times\Lambda\). Let \(J := \mathrm{D}_\mathbf{x}\boldsymbol{\Phi}(\mathbf{x}_0,\lambda_0)\) and let \(q := \dim(\ker(J)) > 0\). For \(\mathbf{x}\in X\), we denote its direct sum decomposition into \(\ker(J)\oplus\ker(J)^\perp\) by
\begin{equation}
    \mathbf{x} = \mathbf{x}^\parallel+\mathbf{x}^\perp\qquad(\mathbf{x}^\parallel,\mathbf{x}^\perp)\in\ker(J)\times\ker(J)^{\perp} \label{def:decompose}
\end{equation}
Let \(\mathbf{x}_0 = \mathbf{x}^\parallel_0+\mathbf{x}^\perp_0\). Then for neighbourhood \({\mathcal{N}}(\mathbf{x}^\parallel_0,\lambda_0)\), there exists a neighbourhood \({\mathcal{N}}(\mathbf{x}^\perp_0)\) along with a \(q\)-dimensional reduced map \(\phi:{\mathcal{N}}(\mathbf{x}^\parallel_0,\lambda_0)\to{\mathcal{N}}(\mathbf{x}^\perp_0)\) where the zeroes of \(\phi\) are in one-to-one correspondence with those of \(\boldsymbol{\Phi}\), i.e., \(\forall\,(\mathbf{x}^\parallel,\lambda)\in{\mathcal{N}}(\mathbf{x}^\parallel_0,\lambda_0)\)
\begin{equation}
    \boldsymbol{\Phi}(\mathbf{x}^\parallel+\mathbf{x}^\perp,\lambda) = 0\iff\phi(\mathbf{x}^\parallel,\lambda) = 0 \label{eq:equivalence}
\end{equation}

The reduced map \(\phi\) in Theorem \ref{thm:LSR} is in general not unique. Its main insight is preservation of topological features of the sets of zeros of the original high-dimensional map \(\boldsymbol{\Phi}\). We illustrate this relationship graphically in Figure \ref{fig:pitchfork}. In the Figure, \(\{\phi(\mathbf{x}^\parallel,\lambda) = 0\}\), which lies in \({\mathcal{N}}(\mathbf{x}^\parallel_0,\lambda_0)\subset\ker(J)\times\R\), is identified with the zero set of \(\boldsymbol{\Phi}\), which lies in \(\{P\boldsymbol{\Phi}(\mathbf{x}^\parallel,\lambda) = 0\}\), where \(P\) is a linear projection onto \(\ker(J)\). The goal of this paper is to derive an estimate for the region \({\mathcal{N}}(\mathbf{x}^\parallel_0,\lambda_0)\times{\mathcal{N}}(\mathbf{x}^\perp_0)\) in which the reduced map \eqref{eq:equivalence} obtained from {Lyapunov-Schmidt} reduction is valid.

\section{Bounds on Lyapunov-Schmidt reduction \label{sec:bounds}}
In this section, we  establish estimates for the bounds of validity for Lyapunov-Schmidt reduction for finite-dimensional systems of nonlinear algebraic equations.
We consider a smooth map \(\boldsymbol{\Phi}\) from Theorem \ref{thm:LSR} with Jacobian \(J = \mathrm{D}_\mathbf{x}\boldsymbol{\Phi}(\mathbf{x}_0,\lambda_0)\) and define a linear projection operator \(P\)
\begin{equation}
    P:\R[n]\to\mathrm{range}(J)\label{def:project}
\end{equation}
and two direct sum decompositions of \(\R[n]\)
\begin{equation}
    \R[n] = \ker(J)\oplus\ker(J)^{\perp} = \mathrm{range}(J)^\perp\oplus\mathrm{range}(J) \label{eq:orth_split}
\end{equation}
{Note that while \(P\) is not unique, its choice does not influence the derived bounds since it does not appear in the calculations.} Recall that subspaces \(\ker(J)^{\perp} = \mathrm{range}(J)\) and analogously \(\mathrm{range}(J)^{\perp} = \ker(J)\), which is often convenient in carrying out the calculations. Throughout the text we will split the state into components along \(\ker(J)\) and \(\ker(J)^{\perp}\) as \(\mathbf{x} = \mathbf{x}^{\parallel} + \mathbf{x}^{\perp}\), which we have already introduced in \eqref{def:decompose}.  We always assume \(\ker(J)\) is nonempty.

Following \cite[Chapter VII]{Golubitsky1985}, we draw the following equivalence:
\begin{equation}
    \boldsymbol{\Phi}(\mathbf{x},\lambda) = 0\iff
    \begin{cases}
        P\boldsymbol{\Phi}(\mathbf{x},\lambda)      = 0 & \\
        (I-P)\boldsymbol{\Phi}(\mathbf{x},\lambda)  = 0 &
    \end{cases}\label{eq:equiv}
\end{equation}
The splitting \eqref{eq:equiv} is a key step in the Lyapunov-Schmidt reduction procedure. The important observation from this splitting is that the ImFT can be applied to the first equation in \eqref{eq:equiv} once we observe that the map \(P\boldsymbol{\Phi}\) restricted to \(\mathrm{range}(J)\) is invertible. This fact can be used to find an implicit map from \(\mathbf{x}^{\perp}\) to \(\mathbf{x}^{\parallel}\), which can in turn be used to eliminate the dependence on \(\mathbf{x}^{\perp}\) in the second equation in \eqref{eq:equiv}, which defines a system of lower state dimension whose solutions are in one-to-one correspondence with the original system. Application of ImFT to solve \(P\boldsymbol{\Phi}(\mathbf{x}^{\parallel}+\mathbf{x}^{\perp},\lambda) = 0\) is also what makes the resulting reduction hold only locally. In the remainder of this section we apply the bounds stated in Theorem \ref{thm:ImFT} to this step, which in turn establishes an estimate for the region of validity of the equivalence \eqref{eq:equivalence} between the original system and the recovered low-dimensional system.

To state the required bounds, we must compute the quantities \(M_x, M_y, \mathbf{L}_x(r_x), \mathbf{L}_{\mathbf{y}}(r_x,r_y)\) in the framework of Theorem \ref{thm:imftbounds}. To do this, it is first helpful to introduce some additional notation. First, we define a  map \(\Xi:\ker(J) \times \ker(J)^{\perp} \times \R[m] \) \begin{equation}
    \Xi(\mathbf{x}^\parallel,\mathbf{x}^\perp,\lambda) := \boldsymbol{\Phi}(\mathbf{x}^\parallel+\mathbf{x}^\perp,\lambda).\label{def:Xi}
\end{equation}
Since \(\dim(\ker(J)) = q\) and \(\dim(\mathrm{range}(J)) = n-q\), we can define an orthonormal basis \(\{\mathbf{v}_1, \dots, \mathbf{v}_q\}\) that spans \(\ker(J)\) and an orthonormal basis \(\{\mathbf{v}_1^{\perp}, \dots, \mathbf{v}_{n-q}^{\perp}\}\) that spans \(\ker(J)^{\perp}\). We then define the matrices \(V\in\R[n \times q]\) and \(V^{\perp}\in\R[n\times(n-q)]\) as follows:
\begin{equation}
    V = \begin{bmatrix} \mathbf{v}_1 & \dots & \mathbf{v}_q \end{bmatrix}, \ \
    V^{\perp} = \begin{bmatrix} \mathbf{v}_1^{\perp} & \dots & \mathbf{v}_{n-q}^{\perp} \end{bmatrix}. \label{def:V}
\end{equation}
In the following Lemma, we derive expressions for the Jacobian matrices of partial derivatives of \(\Xi(\mathbf{x}^{\parallel}, \mathbf{x}^{\perp}, \lambda)\), which will play an important role in the upcoming calculations.

\lem \label{lem:Jacobians}
For \eqref{def:Xi}, the matrices of partial derivatives with respect to \((\mathbf{x}^{\parallel},\lambda)\) and to \(\mathbf{x}^{\perp}\) read
\begin{equation}
    \mathrm{D}_{(\mathbf{x}^{\parallel}, \lambda)}\Xi = \begin{bmatrix}\mathrm{D}_{\mathbf{x}}\boldsymbol{\Phi}(\mathbf{x},\lambda)V&\mathrm{D}_{\lambda}\boldsymbol{\Phi}(\mathbf{x},\lambda)\end{bmatrix}, \label{eq:xi_x}
\end{equation}
\begin{equation}
    \mathrm{D}_{\mathbf{x}^{\perp}}\Xi = \mathrm{D}_{\mathbf{x}} \boldsymbol{\Phi}(\mathbf{x},\lambda) V^{\perp}. \label{eq:xi_y}
\end{equation}
\begin{proof}
    From standard linear algebra, we can know that we can write {\(\mathbf{x} = \Gamma(\boldsymbol{\alpha},\boldsymbol{\beta}) = V\boldsymbol{\alpha} + V^{\perp}\boldsymbol{\beta}\) for some \(\boldsymbol{\alpha}\in\R[q]\) and \(\boldsymbol{\beta}\in\R[n-q]\).} Applying the chain rule gives us \(\mathrm{D}_{\mathbf{x}^{\parallel}} \Xi(\mathbf{x}^{\parallel}, \mathbf{x}^{\perp},\lambda) = \mathrm{D}_{\mathbf{x}} \mathbf{\Phi}(\mathbf{x},\lambda) \mathrm{D}_{\boldsymbol{\alpha}}\Gamma(\boldsymbol{\alpha},\boldsymbol{\beta})\) and \eqref{eq:xi_x} follows. Similar application of chain rule implies \eqref{eq:xi_y}
\end{proof}

Let \(\{\mathbf{w}_1, \dots, \mathbf{w}_{n-q}\}\) be an orthonormal basis of \(\mathrm{range}(J)\), and define the matrix \(W\in\R[n \times (n-q)]\) as
\begin{equation}
    W = \begin{bmatrix} \mathbf{w}_1, \dots, \mathbf{w}_{n-q}\end{bmatrix} \label{def:W}
\end{equation}
We can equivalently express the first of the two systems in the splitting \eqref{eq:equiv} as an \((n-q)\)-dimensional system of equations along \(\mathrm{range}(J)\),
\begin{equation}
    W^\top\mathbf{\Phi}(\mathbf{x},\lambda) = W^\top\Xi(\mathbf{x}^{\parallel},\mathbf{x}^{\perp},\lambda) = 0. \label{eq:sys_for_reduction}
\end{equation}
This is the system for which we compute the ImFT bounds, utilising the derivative matrix definitions of Lemma \ref{lem:Jacobians}. We are now ready to compute the relevant quantities from Theorem \ref{thm:imftbounds}, and applying Lemma \ref{lem:Jacobians}
\begin{itemize}
    \item First, the quantity \(M_x :=  M_{(\mathbf{x}^\parallel,\lambda)}\) reads
    \begin{align}
        M_{(\mathbf{x}^\parallel,\lambda)}  = & \norm{\mathrm{D}_{(\mathbf{x}^\parallel,\lambda)}W^\top\Xi(\mathbf{x}^\parallel_0,\mathbf{x}^\perp_0,\lambda_0)} \notag     \\                      
        =                                     & \norm{\begin{bmatrix}W^\top JV&W^\top\mathrm{D}_\lambda\boldsymbol{\Phi}(\mathbf{x}_0,\lambda_0)\end{bmatrix}} \notag \\
        =                                     & \boxed{\norm{\begin{bmatrix}0 &W^\top\mathrm{D}_\lambda\boldsymbol{\Phi}(\mathbf{x}_0,\lambda_0)\end{bmatrix}}}
        \label{def:M_x-LSR}
    \end{align} where we used the fact that \(JV = 0\) by definition, since the columns of \(V\) are in \(\ker(J)\).
    \item Next, the quantity \(M_y := M_{\mathbf{x}^{\perp}}\) reads
    \begin{equation}
        M_{\mathbf{x}^\perp} = \norm{\big(\mathrm{D}_{\mathbf{x}^\perp}\Xi(\mathbf{x}^\parallel_0,\mathbf{x}^\perp_0,\lambda_0)\big)^{-1}} = \boxed{\norm{(W^\top J V^{\perp})^{-1}}}.\label{def:M_y-LSR}
    \end{equation}
    \item Finally, to compute \(\mathbf{L}_{\mathbf{x}}(r_x) := \mathbf{L}_{(\mathbf{x}^\parallel,\lambda)}(r^\parallel)\), note that
    \[\begin{aligned}
        &\xi_1 (\mathbf{x}^{\parallel},\lambda)  :=  \,\mathrm{D}_{(\mathbf{x}_{\parallel},\lambda)}W^\top\Xi(\mathbf{x}^{\parallel},\mathbf{x}^{\perp}_0,\lambda) \\
                 & - \mathrm{D}_{(\mathbf{x}^\parallel,\lambda)}W^\top\Xi(\mathbf{x}^\parallel_0,\mathbf{x}^\perp_0,\lambda_0) \\
               = & \begin{bmatrix}W^\top\mathrm{D}_{\mathbf{x}} \boldsymbol{\Phi}(\mathbf{x}^{\parallel}+\mathbf{x}^\perp_0,\lambda)V&W^\top\mathrm{D}_{\lambda}\boldsymbol{\Phi}(\mathbf{x}^{\parallel}+\mathbf{x}^\perp_0,\lambda)\end{bmatrix} \\
                 & - \begin{bmatrix} 0 & W^\top\mathrm{D}_{\lambda}\boldsymbol{\Phi}(\mathbf{x}_0,\lambda_0)\end{bmatrix}.
    \end{aligned}\]
    \item Then we can define the quantity
    \begin{gather}
        \mathbf{L}_{(\mathbf{x}^\parallel,\lambda)}(r^\parallel) := \sup\Big\{\norm{\xi_1}:(\mathbf{x}^\parallel,\lambda)\in\mathcal{B}\big((\mathbf{x}^\parallel_0,\lambda_0),r^\parallel\big)\Big\}.\label{def:L_x-LSR}
    \end{gather}
    \item Analogously, we can define
    \[\begin{aligned}
        \xi_2(\mathbf{x}^{\parallel}, \mathbf{x}^{\perp},\lambda) & = \mathrm{D}_{\mathbf{x}^\perp} W^\top\Xi(\mathbf{x}^\parallel,\mathbf{x}^\perp,\lambda) \\ 
                                                                  & \qquad - \mathrm{D}_{\mathbf{x}^\perp} W^\top\Xi(\mathbf{x}^\parallel_0,\mathbf{x}^\perp_0,  \lambda_0) \\
                                                                  & = W^\top\mathrm{D}_{\mathbf{x}}\mathbf{\Phi}(\mathbf{x},\lambda)V^{\perp}-W^\top JV^{\perp}
    \end{aligned}\]
    \item Using the above, \(\mathbf{L}_{\mathbf{y}}(r_x, r_y) := \mathbf{L}_{\mathbf{x}^\perp}(r^\parallel,r^\perp)\) reads,
    \begin{multline}
        \mathbf{L}_{\mathbf{x}^\perp}(r^\parallel,r^\perp) := \sup\Big\{\norm{\xi_2(\mathbf{x}^{\parallel}, \mathbf{x}^{\perp},\lambda)}:      \\
        \big((\mathbf{x}^\parallel,\lambda),\mathbf{x}^\perp\big) \in\mathcal{B}\big((\mathbf{x}^\parallel_0,\lambda_0),r^\parallel\big)\times\mathcal{B}\big(\mathbf{x}^\perp_0,r^\perp\big)\Big\}. \label{def:L_y-LSR}
    \end{multline}
\end{itemize}

We are now ready to state our main result.

\begin{figure}[ht!]
    \centering
    \includegraphics[width=\linewidth]{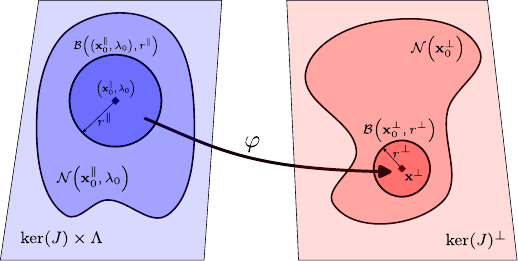}
    \caption{sketch of map \(\varphi:\mathcal{B}\big((\mathbf{x}^\parallel_0,\lambda_0),r^\parallel\big)\to\mathcal{B}\big(\mathbf{x}^\perp_0,r^\perp\big)\) \label{fig:map}}
    \vspace{-6mm}
\end{figure}

\thm\label{thm:LSRbounds} Let \(X\subset\R[n]\) and \(\Lambda\subset\R[m]\) be open sets. Let \(\boldsymbol{\Phi}:X\times\Lambda\to\R[n]\) be a smooth map with \(\boldsymbol{\Phi}(\mathbf{x}_0,\lambda_0) = 0\) for some \((\mathbf{x}_0,\lambda_0)\in X\times\Lambda\). Let \(J := \mathrm{D}_\mathbf{x}\boldsymbol{\Phi}(\mathbf{x}_0,\lambda_0)\) and let \(q := \dim(\ker(J)) > 0\). For any \(r^\parallel,r^\perp>0\) satisfying
\[\begin{aligned}
    \mathbf{L}_{(\mathbf{x}^\parallel,\lambda)}(r^\parallel)r^\parallel+\mathbf{L}_{\mathbf{x}^\perp}(r^\parallel,r^\perp)r^\perp & < \frac{r^\perp}{M_{\mathbf{x}^\perp}}-M_{(\mathbf{x}^\parallel,\lambda)}r^\parallel \\
    M_{\mathbf{x}^\perp}\mathbf{L}_{\mathbf{x}^\perp}(r^\parallel,r^\perp)                                                        & < 1
\end{aligned}\]
where the relevant quantities are defined by the equations \eqref{def:M_x-LSR}, \eqref{def:M_y-LSR}, \eqref{def:L_x-LSR}, \eqref{def:L_y-LSR}, there exists a smooth implicit map \(\varphi:\mathcal{B}\big((\mathbf{x}^\parallel_0,\lambda_0),r^\parallel\big)\to\mathcal{B}\big(\mathbf{x}^\perp_0,r^\perp\big)\)
satisfying
\[
    \Xi\big(\mathbf{x}^\parallel,\varphi(\mathbf{x}^\parallel,\lambda),\lambda\big) = \boldsymbol{\Phi}\big(\mathbf{x}^\parallel+\varphi(\mathbf{x}^\parallel,\lambda),\lambda\big) = 0.
\]for all \((\mathbf{x}^\parallel,\lambda)\in\mathcal{B}\big((\mathbf{x}^\parallel_0,\lambda_0),r^\parallel\big)\)

\begin{proof}
    The above follows directly from Theorem \ref{thm:LSR} and from the bounds on the Implicit Function Theorem stated in Theorem \ref{thm:imftbounds}.
\end{proof}

In Fig. \ref{fig:map} we provide a graphical representation of the implicit map \(\varphi\), {its estimated open-ball \((\mathcal{B})\) bounds of validity established in Theorem \ref{thm:LSRbounds} which are subsets of the maximal neighbourhoods of validity of the implicit map \((\mathcal{N})\)}. For the sake of completion, we state the final steps in the procedure of {Lyapunov-Schmidt} reduction. The reduced \(q\)-dimensional map obtained by plugging the implicit map \(\varphi(\mathbf{x}^{\parallel},\lambda)\) into the second set of equations in \eqref{eq:equiv} as
\[
    \phi(\mathbf{x}^\parallel,\lambda) := (I-P)\boldsymbol{\Phi}\big(\mathbf{x}^\parallel+\varphi(\mathbf{x}^\parallel,\lambda),\lambda\big)
\]
satisfying \(\phi(\mathbf{x}^\parallel,\lambda) = 0\iff\boldsymbol{\Phi}\big(\mathbf{x}^\parallel+\varphi(\mathbf{x}^\parallel,\lambda),\lambda\big) = 0\). To obtain a true \(q\)-dimensional reduced map we can express \(\phi\) in terms of coordinates along \(\mathrm{range}(J)^{\perp}\) to complete the reduction. {We also note that although these bounds are not expected to be maximal in general, the following section will showcase an example for which they are.}

\section{Example: recurrent neural network model \label{sec:example}}

In this section, we develop in detail the computation of the bounds established in Theorem \ref{thm:LSRbounds} on an illustrative two-dimensional example.

\subsection{System}
For \(\R[2]\ni\mathbf{x} := \begin{bmatrix}x_1\\x_2\end{bmatrix}\) and \(\lambda\in\R\), consider the following two-node dynamic neural network model described by:
\begin{equation}
    \dot{\mathbf{x}} = \boldsymbol{\Phi}(\mathbf{x},\lambda)
                     = -\mathbf{x}+\tanh\left(\begin{bmatrix}0&\lambda\\\lambda&0\end{bmatrix}\mathbf{x}\right) \label{eq:example}
\end{equation}
where the parameter \(\lambda\)'s magnitude regulates the slope of the saturation function \(\tanh\), and sign regulates whether the interconnection between the two neural network nodes is mutually excitatory \((\lambda>0)\) or mutually inhibitory \((\lambda<0)\). {Models of this form are commonly encountered across different contexts, for example in the study of recurrent neural networks \cite{cheng2006multistability} and nonlinear opinion dynamics \cite{bizyaeva2022nonlinear}.}

\vspace{-4mm}
\begin{figure}[ht!] 
    \centering
    \begin{subfigure}{.24\textwidth}
        \centering
        \includegraphics[width=\linewidth]{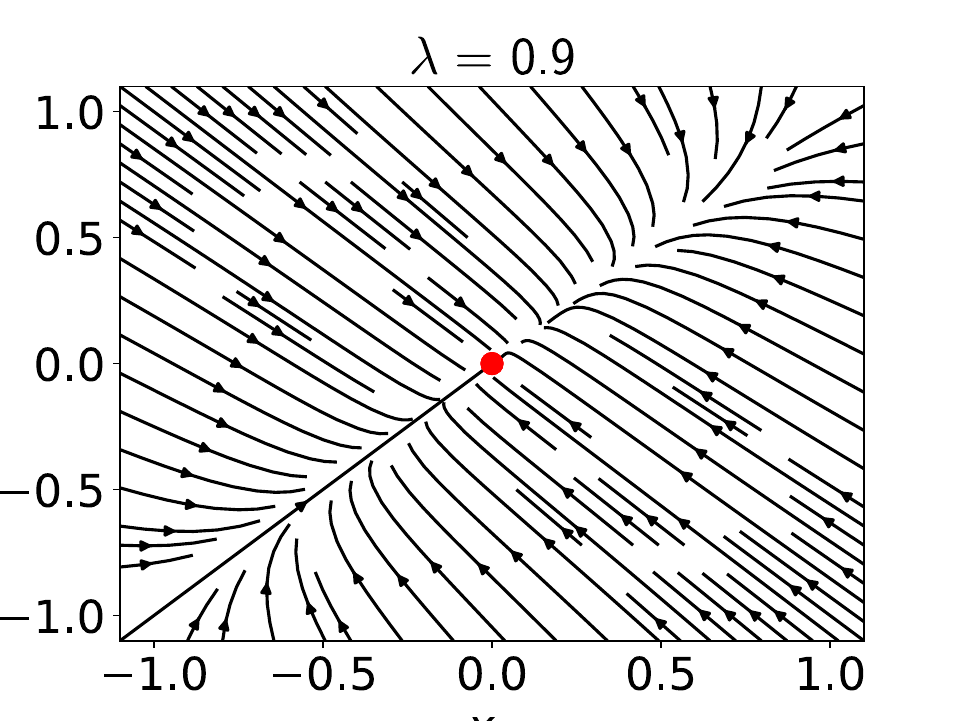}
    \end{subfigure}%
    \begin{subfigure}{.24\textwidth}
        \centering
        \includegraphics[width=\linewidth]{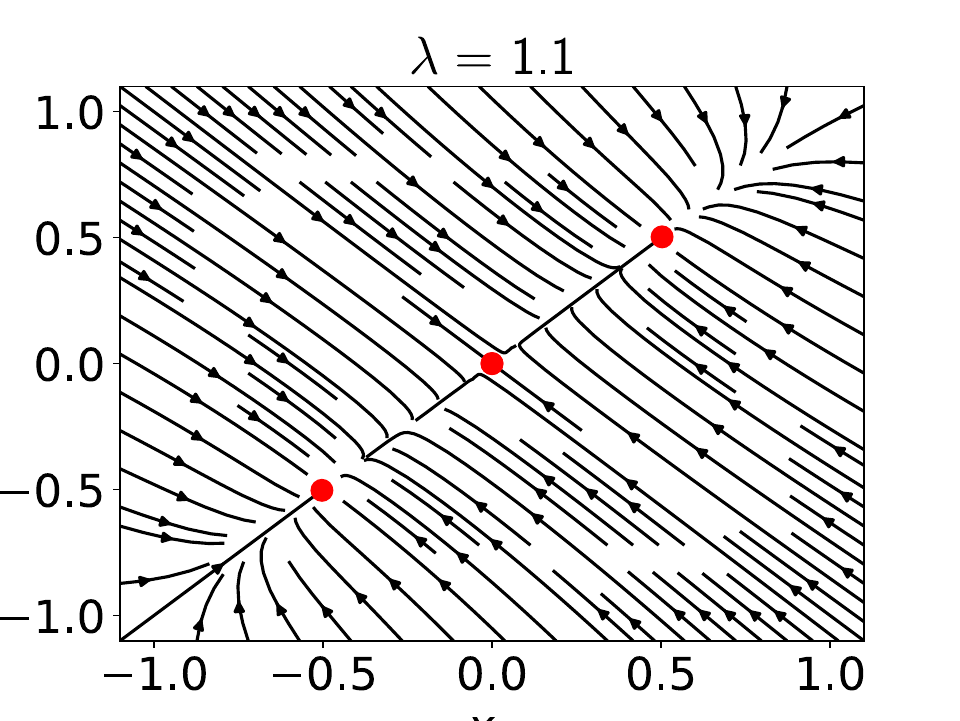}
    \end{subfigure}
    \caption{Flow of \eqref{eq:example} close to the pitchfork bifurcation}\label{fig:flow}
\end{figure}
\vspace{-4mm}

At \(\lambda = 1\) the system undergoes a pitchfork bifurcation, with a transition between a single globally stable equilibrium point at the origin \((x_1,x_2) = (0,0)\) to an unstable equilibrium point at \((0,0)\) and two simultaneously stable equilibria at \(\pm(x^\star,x^\star)\) where \(x^\star\) is implicitly defined through the equation \(\tanh(\lambda\tanh(\lambda x^\star)) = x^\star\) - see Fig. \ref{fig:flow}.

Now, \(\boldsymbol{\Phi}(\mathbf{0},\lambda) = \mathbf{0}\implies\mathbf{0}\) is an equilibrium \(\forall\,\lambda\in\R\).
\[\begin{aligned}
        \mathrm{D}_\mathbf{x}\boldsymbol{\Phi}(\mathbf{x},\lambda) & = \begin{bmatrix}-1&\lambda\sech^2(\lambda x_2)\\\lambda\sech^2(\lambda x_1)&-1\end{bmatrix} \\
        \implies J                                                 & = \mathrm{D}_\mathbf{x}\boldsymbol{\Phi}(\mathbf{0},\lambda) = \begin{bmatrix}-1&\lambda\\\lambda&-1\end{bmatrix}
    \end{aligned}\]
Hence, \(J\) is singular when \(\lambda\in\{-1,1\}\). We have determined the bounds only for \(\lambda=1\)since \(\lambda=-1\) follows analogously.

\vspace{-1mm}
\subsection{Computation of Bounds}
We shall now compute the bounds established in Theorem \ref{thm:LSRbounds} for \eqref{eq:example} in a step-by-step manner.

\begin{enumerate}
    \itemsep 0.5em
    \item Compute the Jacobian \(J = \mathrm{D}_\mathbf{x}\boldsymbol{\Phi}(\mathbf{0},1) = \begin{bmatrix}-1&1\\1&-1\end{bmatrix}\).
    \item To compute the linear projection \(P:\R[2]\to\mathrm{range}(J)\) defined in \eqref{def:project}, {we chose the unique orthogonal projection matrix}, which works out to
    \[
        P = {W(W^\top W)^{-1}W^\top} = \frac12\begin{bmatrix}1&-1\\-1&1\end{bmatrix}\label{eq:P}
    \]
    \item For the orthogonal splitting of \(\R[2]\) as per \eqref{eq:orth_split}, notice that for \(\mathbf{v}_1 = \displaystyle\frac{1}{\sqrt2}\begin{bmatrix}1\\1\end{bmatrix}\) and \(\mathbf{v}_2 = \displaystyle\frac{1}{\sqrt2}\begin{bmatrix}-1\\1\end{bmatrix}\), we have \label{eq:basis}
          \[\begin{aligned}
                \ker(J)           & = \mathrm{span}(\mathbf{v}_1) \\
                \mathrm{range}(J) & = \mathrm{span}(\mathbf{v}_2)
            \end{aligned}\implies\begin{aligned}
                \ker(J)^\perp           & = \mathrm{span}(\mathbf{v}_2) \\
                \mathrm{range}(J)^\perp & = \mathrm{span}(\mathbf{v}_1)
            \end{aligned}\]
          Hence, we obtain \(V\) and \(V^\perp\) as defined as \eqref{def:V} and \(W\) as defined in \eqref{def:W} \(\implies V=\mathbf{v}_1, V^\perp=\mathbf{v}_2, W=\mathbf{v}_2\)
    \item In order to compute \(M_{(\mathbf{x}^\parallel,\lambda)}\) from its definition in \eqref{def:M_x-LSR} we first compute \(\mathrm{D}_{\lambda}\boldsymbol{\Phi}(\mathbf{0},1)\) as follows:
          \[
              \mathrm{D}_{\lambda}\boldsymbol{\Phi}(\mathbf{x},\lambda) = \begin{bmatrix}x_2\sech^2(\lambda x_2)\\x_1\sech^2(\lambda x_1)\end{bmatrix} \implies \mathrm{D}_{\lambda}\boldsymbol{\Phi}(\mathbf{0},1) = \mathbf{0}
          \]
          to finally obtain
          \[\begin{aligned}
                W^\top\mathrm{D}_\lambda\boldsymbol{\Phi}(\mathbf{0},1) & = \frac{1}{\sqrt{2}}\begin{bmatrix}-1&1\end{bmatrix}\begin{bmatrix}0\\0\end{bmatrix} = 0 \\
                \implies M_{(\mathbf{x}^\parallel,\lambda)}                  & = \norm{\begin{bmatrix}0 &W^\top\mathrm{D}_\lambda\boldsymbol{\Phi}(\mathbf{0},1)\end{bmatrix}} = \boxed{0}
            \end{aligned}\]
    \item Now to compute \(M_{\mathbf{x}^\perp}\) as per \eqref{def:M_y-LSR}
          \[\begin{aligned}
                M_{\mathbf{x}^\perp} & = \norm{(W^\top J V^{\perp})^{-1}}                                                                                                   \\
                                     & = \norm{\left(\frac12\begin{bmatrix}-1&1\end{bmatrix}\begin{bmatrix}-1&1\\1&-1\end{bmatrix}\begin{bmatrix}-1\\1\end{bmatrix}\right)^{-1}} \\
                                     & = \norm{\begin{bmatrix}-2\end{bmatrix}^{-1}} = \boxed{\frac12}
            \end{aligned}\]
    \item In order to compute \(\mathbf{L}_{(\mathbf{x}^\parallel,\lambda)}(r^\parallel)\) from its definition in \eqref{def:L_x-LSR}, we first need to calculate
          \[
                \xi_1(\mathbf{x}^{\parallel},\lambda) = \begin{bmatrix}W^\top\mathrm{D}_{\mathbf{x}}\boldsymbol{\Phi}(\mathbf{x}^{\parallel},\lambda)V&W^\top\mathrm{D}_{\lambda}\boldsymbol{\Phi}(\mathbf{x}^{\parallel},\lambda)\end{bmatrix}
          \]
          for which we calculate the individual terms, given that \(\displaystyle\mathbf{x}^{\parallel} = \frac{\gamma}{\sqrt{2}}\begin{bmatrix}1\\1\end{bmatrix}\) for some \(\gamma\in\R\)
          \[\begin{aligned}
                \mathrm{D}_{\mathbf{x}}\boldsymbol{\Phi}\left(\mathbf{x}^{\parallel},\lambda\right) & = \begin{bmatrix}-1&\lambda\sech^2\left(\lambda\frac{\gamma}{\sqrt{2}}\right)\\\lambda\sech^2\left(\lambda\frac{\gamma}{\sqrt{2}}\right)&-1\end{bmatrix} \\
                \implies & W^\top\mathrm{D}_{\mathbf{x}}\boldsymbol{\Phi}\left(\mathbf{x}^{\parallel},\lambda\right)V = 0 \\
                \mathrm{D}_{\lambda}\boldsymbol{\Phi}\left(\mathbf{x}^{\parallel},\lambda\right) & = \frac{\gamma}{\sqrt{2}}\sech^2\left(\frac{\lambda\gamma}{\sqrt{2}}\right)\begin{bmatrix}1\\1\end{bmatrix} \\
                \implies & W^\top\mathrm{D}_{\lambda}\boldsymbol{\Phi}\left(\mathbf{x}^{\parallel},\lambda\right) = 0
            \end{aligned}\]
          yielding \(\xi_1(\mathbf{x}^{\parallel},\lambda) = \begin{bmatrix}0&0\end{bmatrix}\implies\boxed{\mathbf{L}_{(\mathbf{x}^\parallel,\lambda)}(r^\parallel) = 0}\)
    \item In order to compute \(\mathbf{L}_{\mathbf{x}^\perp}(r^\parallel,r^\perp)\) as defined in \eqref{def:L_y-LSR}, we first need to compute
          \[
              \xi_2(\mathbf{x}^{\parallel}, \mathbf{x}^{\perp},\lambda) = W^\top\mathrm{D}_{\mathbf{x}} \mathbf{\Phi}(\mathbf{x},\lambda) V^{\perp} - W^\top J V^{\perp}
          \]
          for which we calculate the individual terms
          \[\small \begin{aligned}
                & W^\top\mathrm{D}_{\mathbf{x}} \mathbf{\Phi}(\mathbf{x},\lambda) V^{\perp} = \frac12\begin{bmatrix}-1&1\end{bmatrix}\begin{bmatrix}1+\lambda\sech^2(\lambda x_2)\\-1-\lambda\sech^2(\lambda x_1)\end{bmatrix} \\
                & \qquad\qquad\quad\;\;\; = \begin{bmatrix}-1-\frac{\lambda}{2}(\sech^2(\lambda x_1)+\sech^2(\lambda x_2))\end{bmatrix} \\
                & W^\top J V^{\perp} = \frac12\begin{bmatrix}-1&1\end{bmatrix}\begin{bmatrix}-1&+1\\+1&-1\end{bmatrix}\begin{bmatrix}-1\\1\end{bmatrix} = \begin{bmatrix}-2\end{bmatrix} \\
                & \boxed{\xi_2(\mathbf{x}^{\parallel}, \mathbf{x}^{\perp},\lambda) = \begin{bmatrix}1-\frac{\lambda}{2}\big(\sech^2(\lambda x_1)+\sech^2(\lambda x_2)\big)\end{bmatrix}}
            \end{aligned}\]
          Hence, for \(\big((\mathbf{x}^\parallel,\lambda),\mathbf{x}^\perp\big)\in\mathcal{B}\big((\mathbf{0},1),r^\parallel\big)\times\mathcal{B}\big(\mathbf{0},r^\perp\big)\)
          \[\begin{aligned}
                  & \mathbf{L}_{\mathbf{x}^\perp}(r^\parallel,r^\perp) := \sup\Big\{\norm{\xi_2(\mathbf{x}^{\parallel}, \mathbf{x}^{\perp},\lambda)}\Big\} \\
                  & = \sup\left\{1-\frac{\lambda}{2}\big(\sech^2(\lambda x_1)+\sech^2(\lambda x_2)\big)\right\} \\
                  & = 1 - \frac12\inf\left\{\lambda\big(\sech^2(\lambda x_1)+\sech^2(\lambda x_2)\big)\right\} \\
                  & = \boxed{1 - \min\{0,\lambda\}}
            \end{aligned}\]
\end{enumerate}

We have all the requisite quantities ready to finally compute the bounds. The second constraint in Theorem \ref{thm:LSRbounds} simplifies to the statement \(0 < 1\) and is therefore satisfied for any choice of \(r^{\parallel}\) and \(r^{\perp}\). The second constraint yields the  condition
\begin{equation*}
    \mathbf{L}_{\mathbf{x}^\perp}(r^\parallel,r^\perp) = 1 - \min\{0,\lambda\} < 2 \implies \boxed{\min\{0,\lambda\} > -1}
\end{equation*}
which is satisfied for any \(r^{\perp} > 0\) provided that \(r^{\parallel} < 2\). Hence, according to Theorem \ref{thm:LSRbounds}, for any \(\boxed{r^{\perp} > 0}\) and \(\boxed{0 < r^{\parallel} < 2}\)
there exists a smooth implicit map
\[
    \varphi:\mathcal{B}\big((0,1),r^\parallel\big)\to\mathcal{B}(0,r^\perp)
\]
satisfying
\[
    \boxed{\varphi(\alpha\mathbf{v}_{1},\lambda)+\frac12\begin{bmatrix}1&-1\\-1&1\end{bmatrix}\tanh\Big(\lambda\big(\alpha\mathbf{v}_1-\varphi(\alpha \mathbf{v}_{1},\lambda)\big)\Big) = 0}.
\]
Note that these bounds are interpretable based on our understanding of the bifurcation diagram of the simple model.  Recall that there is a subcritical pitchfork bifurcation at \(\lambda = -1\), with new branches of equilibria appearing for values of \(\lambda < -1\). As long as \(r^{\parallel}\) is less than 2, the open ball \(\mathcal{B}\big((0,1), r^{\perp}\big)\) does not intersect with this second bifurcation point. This means that for this particular example, the bounds we arrived at are as tight as possible.

\subsection{Completion of LS reduction}
Note that the reduced map \(\phi\) is obtained from the second equivalence condition in \eqref{eq:equiv} as follows:
\[\begin{aligned}
    (I-P)\boldsymbol{\Phi}(\mathbf{x},\lambda) & = \frac12\begin{bmatrix}1&1\\1&1\end{bmatrix}\left(-\mathbf{x}+\tanh\left(\begin{bmatrix}0&\lambda\\\lambda&0\end{bmatrix}\mathbf{x}\right)\right) \\
                                               & = -\alpha\mathbf{v}_1+\frac12\begin{bmatrix}1&1\\1&1\end{bmatrix}\tanh\big(\lambda(\alpha\mathbf{v}_1-\beta\mathbf{v}_2)\big)
\end{aligned}\]
Substituting \(\varphi\) into above, we obtain:
\[
    \boxed{\phi(\alpha\mathbf{v}_1,\lambda) = -\alpha\mathbf{v}_1+\frac12\begin{bmatrix}1&1\\1&1\end{bmatrix}\tanh\Big[\lambda\big(\alpha\mathbf{v}_1-\varphi(\alpha\mathbf{v}_{1},\lambda)\big)\Big]}
\]
To express this mapping as a 1-dimensional equation we project \(\phi\) onto \( \mathrm{range}(J)^\perp = \ker(J^\top) = \mathrm{span}(\mathbf{v}_1)\), obtaining our final reduced equation
\[\begin{aligned}
    g(\alpha,\lambda) & = \langle\mathbf{v}_1,\phi(\alpha\mathbf{v}_1,\lambda)\rangle                                                       \\
                 & = -\alpha+\mathbf{v}_1^\top\tanh\Big[\lambda\big(\alpha\mathbf{v}_1-\varphi(\alpha\mathbf{v}_{1},\lambda)\big)\Big]
\end{aligned}\]
The above 1D equation is a reduced-order model for the bifurcation diagram of the original system, obtained via Lyapunov-Schmidt reduction. Since the mapping \(\varphi\) is not explicitly known, in practice we compute derivatives of the above reduced system to find a series expansion for it in order to study its bifurcations;for example see \cite[II.9]{Golubitsky1985}.




\section{Conclusions \label{sec:conclusion}}
\label{cnclsn}
In this article, we derive estimates that bound the validity of Lyapunov-Schmidt reduction, a dimensionality reduction technique frequently used in the analysis of bifurcations in dynamic systems. In order to come up with these estimates, we utilise results on the quantitative estimates on validity bounds for the Implicit Function Theorem, recently derived in \cite{jindal2023imft}. The derived bounds are dependent on the first order derivatives evaluated at the critical point and the bounds on the first order derivatives computed over a region of interest containing said critical point. We also demonstrate a detailed derivation of this bound for a dynamical neural network model that exhibits a pitchfork bifurcation. In future work, the bounds established in Theorem \ref{thm:LSRbounds} can be applied in the context of bifurcation problems for various high-dimensional dynamical systems. For example, it would be interesting to consider how graph structure affects the tightness of the established bounds in bifurcation problems defined over dynamic networks, such as the opinion dynamics models in \cite{bizyaeva2022nonlinear,bizyaeva2023thesis}. It would also be interesting to extend these bounds beyond finite-dimensional systems, since the Lyapunov-Schmidt reduction procedure is defined more generally for Fredholm operators over Banach spaces.


%
\bibliographystyle{IEEEtran}
\bibliography{ref}

\end{document}